\pdfoutput=1
\documentclass[preprint,preprintnumbers,amsmath,amssymb,floatfix,endfloats*]{revtex4}

\usepackage{graphicx}
\usepackage{dcolumn}
\usepackage{bm}
\usepackage{amsfonts}

\DeclareMathAlphabet{\mathsfsl}{OT1}{cmr}{bx}{it}
\begin{document}
\title{Shear band formation in amorphous materials under oscillatory shear deformation}
\author{Nikolai V. Priezjev}
\affiliation{Department of Mechanical and Materials Engineering,
Wright State University, Dayton, OH 45435}
\date{\today}
\begin{abstract}

The effect of periodic shear on strain localization in disordered
solids is investigated using molecular dynamics simulations.  We
consider a binary mixture of one million atoms annealed to a low
temperature with different cooling rates and then subjected to
oscillatory shear deformation with a strain amplitude slightly above
the critical value. It is found that the yielding transition occurs
during one cycle but the accumulation of irreversible displacements
and initiation of the shear band proceed over larger number of
cycles for more slowly annealed glasses. The spatial distribution
and correlation function of nonaffine displacements reveal that
their collective dynamics changes from homogeneously distributed
small clusters to a system-spanning shear band. The analysis of
spatially averaged profiles of nonaffine displacements indicates
that the location of a shear band in periodically loaded glasses can
be identified at least several cycles before yielding. These
insights are important for development of novel processing methods
and prediction of the fatigue lifetime of metallic glasses.

\vskip 0.5in

Keywords: metallic glasses, periodic deformation, yield stress,
molecular dynamics simulations

\end{abstract}

\maketitle

\section{Introduction}

Understanding the relationship between atomic structure and elastic
and plastic deformation properties of metallic glasses and other
disordered solids is important for numerous structural and
biomedical applications~\cite{Hufnagel16,ZhengBio16}. One major
drawback that prevents the widespread use of metallic glasses is the
formation of narrow, extended regions where strain becomes
localized, the so-called shear bands, which lead to structural
failure of the material~\cite{Parisi17,Wolynes17}.  It has long been
recognized that localized plastic events in deformed amorphous
materials involve collective rearrangements of small clusters of
atoms, or shear transformations~\cite{Spaepen77,Argon79}.  In recent
years, the processes of shear band initiation and propagation have
been extensively studied at the atomic level during startup
uniaxial~\cite{ShiFalk05,Ma09,Shin-Pon15,Zuniga18,Feng18} and
shear~\cite{Horbach16,Pastewka19,Priez19star}  deformation with a
constant strain rate.  While the initiation of a shear band
typically occurs at a free surface during tension or compression,
the transition to plastic flow in a sheared periodic domain involves
the formation of a percolating cluster of mobile regions at the
critical strain~\cite{Horbach16}.  Remarkably, it was recently found
that the structure of shear bands in metallic glasses consists of
alternating dilated and densified regions, which originate from the
alignment of Eshelby-like quadrupolar stress
fields~\cite{Zaccone17,Varnik19}. Even though strain localization is
a common phenomenon in amorphous systems, the complete description
of structural processes leading to material failure under imposed
deformation is still missing.

\vskip 0.05in

In the last few years, a number of atomistic simulation studies were
carried out to investigate the failure mechanism and structural
relaxation in amorphous materials subjected to periodic
deformation~\cite{Priezjev13,Sastry13,Reichhardt13,Priezjev14,
Priezjev16,Kawasaki16,Priezjev16a,Sastry17,Priezjev17,OHern17,
GaoSize17,Priezjev18,Priezjev18a,NVP18strload,PriMakrho05,
PriMakrho09,Sastry19band,PriezSHALT20}. Notably, it was found that
rapidly quenched glasses can be mechanically annealed to lower
energy states during a number of subyield cycles, depending on the
strain amplitude and temperature~\cite{Sastry13,Sastry17,
Priezjev18,Priezjev18a}. More recently, it was shown that even lower
energy states can be accessed if the orientation of periodic shear
is alternated every one or several cycles leading to the increase in
strength and shear-modulus anisotropy~\cite{PriezSHALT20}.  On the
other hand, if the strain amplitude is above the critical value, the
yielding transition in well-annealed glasses occurs after a number
of shear cycles, followed by the shear band formation and
stress-strain hysteresis~\cite{Priezjev17}.  It was further
demonstrated that sufficiently large samples under
strain-controlled, tension-compression loading exhibit cyclic
softening and reduced fatigue lifetime~\cite{GaoSize17}. Moreover,
upon cyclic loading after strain localization, the potential energy
within the shear band increases and the particle motion becomes
diffusive, whereas the glass outside the shear band continues
annealing and particle motions are subdiffusive~\cite{Sastry19band}.
However, despite extensive research, the exact mechanism of the
shear band initiation and growth during oscillatory deformation
needs to be further clarified.

\vskip 0.05in

In this paper, the yielding transition and shear band formation in
cyclically deformed binary glasses are studied using molecular
dynamics simulations. The model glass is first annealed well below
the glass transition temperature with different cooling rates, and
then subjected to periodic shear with the strain amplitude greater
than the critical value. It will be shown that while the yielding
transition, marked by the decrease in shear stress and potential
energy amplitudes, takes place during one shear cycle, the fatigue
time is longer for more slowly annealed glasses.  Moreover, the
temporal evolution of spatial distributions and averaged profiles of
nonaffine displacements indicates that the location of a shear band
can be predicted at least several cycles before yielding.

\vskip 0.05in

The reminder of the paper is organized as follows. The next section
contains a description of molecular dynamics simulations and the
deformation procedure.  The time dependence of the potential energy
and shear stress, spatial distribution of nonaffine displacements,
as well as the analysis of the nonaffinity correlation function and
nonaffine displacement profiles are presented in
section\,\ref{sec:Results}.   Brief conclusions are provided in the
last section.

\section{Molecular dynamics (MD) simulations}
\label{sec:MD_Model}

In our study, the model glass is represented by the Kob-Andersen
(KA) binary mixture of two types of atoms, $\alpha,\beta=A,B$, with
strongly non-additive interaction between different types of atoms,
which prevents formation of the crystalline phase~\cite{KobAnd95}.
The parametrization of the KA binary mixture is similar to the model
of the amorphous metal-metalloid alloy $\text{Ni}_{80}\text{P}_{20}$
studied by Weber and Stillinger~\cite{Weber85}. Here, we consider a
relatively large system that consists of $800\,000$ atoms of type
$A$ and $200\,000$ atoms of type $B$, with the total number of atoms
equal to $10^6$. These atoms interact via the pairwise Lennard-Jones
(LJ) potential, as follows:
\begin{equation}
V_{\alpha\beta}(r)=4\,\varepsilon_{\alpha\beta}\,\Big[\Big(\frac{\sigma_{\alpha\beta}}{r}\Big)^{12}\!-
\Big(\frac{\sigma_{\alpha\beta}}{r}\Big)^{6}\,\Big],
\label{Eq:LJ_KA}
\end{equation}
with the standard parametrization $\varepsilon_{AA}=1.0$,
$\varepsilon_{AB}=1.5$, $\varepsilon_{BB}=0.5$, $\sigma_{AA}=1.0$,
$\sigma_{AB}=0.8$, $\sigma_{BB}=0.88$, and
$m_{A}=m_{B}$~\cite{KobAnd95}.   In order to speed up computation of
the interaction forces, the LJ potential was truncated at the cutoff
radius $r_{c,\,\alpha\beta}=2.5\,\sigma_{\alpha\beta}$. For
convenience, the simulation results are presented in the LJ units of
length, mass, energy, and time: $\sigma=\sigma_{AA}$, $m=m_{A}$,
$\varepsilon=\varepsilon_{AA}$, and, consequently,
$\tau=\sigma\sqrt{m/\varepsilon}$. The simulations were carried out
using the LAMMPS parallel code~\cite{Lammps,Stegailov11}, where the
equations of motion were solved numerically using the
velocity-Verlet algorithm with the time step $\triangle
t_{MD}=0.005\,\tau$~\cite{Allen87}.

\vskip 0.05in


The binary mixture was first equilibrated in the liquid state at the
temperature $T_{LJ}=1.0\,\varepsilon/k_B$, where $k_B$ is the
Boltzmann constant. The system temperature was regulated via the
Nos\'{e}-Hoover thermostat~\cite{Allen87,Lammps}. The MD simulations
were conducted in a periodic box at constant volume and the
corresponding density $\rho=\rho_A+\rho_B=1.2\,\sigma^{-3}$. It was
previously demonstrated that at this density, the critical
temperature of the KA model is
$T_c=0.435\,\varepsilon/k_B$~\cite{KobAnd95}. After equilibration,
the binary mixture was linearly cooled to the low temperature
$T_{LJ}=0.01\,\varepsilon/k_B$ with the rates
$10^{-2}\varepsilon/k_{B}\tau$, $10^{-3}\varepsilon/k_{B}\tau$,
$10^{-4}\varepsilon/k_{B}\tau$, and $10^{-5}\varepsilon/k_{B}\tau$,
while keeping the volume constant. The linear size of the simulation
box is $94.10\,\sigma$.

\vskip 0.05in


Once the system was brought to the low temperature
$T_{LJ}=0.01\,\varepsilon/k_B$, it was subjected to oscillatory
shear deformation at constant volume, as follows:
\begin{equation}
\gamma(t)=\gamma_0\,\text{sin}(2\pi t/T),
\label{Eq:shear}
\end{equation}
where $T=5000\,\tau$ is the oscillation period and $\gamma_0=0.075$
is the strain amplitude. The periodic strain was imposed along the
$xz$ plane using the Lees-Edwards periodic boundary
conditions~\cite{Allen87,Lammps}. The corresponding frequency of the
applied deformation is $\omega=2\pi/T=1.26\times10^{-3}\,\tau^{-1}$.
As discussed below, the strain amplitude, $\gamma_0=0.075$, was
chosen to be slightly above the critical value for the yielding
transition at the density
$\rho=1.2\,\sigma^{-3}$~\cite{Priezjev17,Sastry17}. For each value
of the cooling rate, the simulations were performed only in one
sample. During oscillatory deformation, the potential energy, stress
components, system dimensions, and atomic configurations were
regularly saved for the post-processing analysis.

\section{Results}
\label{sec:Results}


A number of recent molecular dynamics simulations studies have shown
that disordered solids under cyclic loading either continue
exploring deeper energy states at sufficiently small strain
amplitudes or eventually undergo a transition to plastic flow within
a shear band if the strain amplitude is above a certain
value~\cite{Sastry13,Reichhardt13,Sastry17,Priezjev17,Priezjev18,
Priezjev18a,Sastry19band,PriezSHALT20}.  The precise determination
of the critical strain amplitude, however, is a challenging problem
because the number of cycles required to reach a dynamic steady
state appears to diverge upon approaching the critical
value~\cite{Sastry13,Reichhardt13}.  In addition, the estimate of
the critical amplitude for binary glasses was shown to be system
size~\cite{Sastry13} and
temperature~\cite{Priezjev13,Priezjev18a,Sastry19band} dependent.
Therefore, in the present study, the MD simulations were performed
at the strain amplitude, $\gamma_0=0.075$, which is slightly larger
than the critical value for the KA binary mixture at the density
$\rho=1.2\,\sigma^{-3}$ and temperatures well below the glass
transition
point~\cite{Priezjev17,Sastry17,Priezjev18a,Sastry19band}.

\vskip 0.05in

\begin{figure}[t]
\includegraphics[width=12.0cm,angle=0]{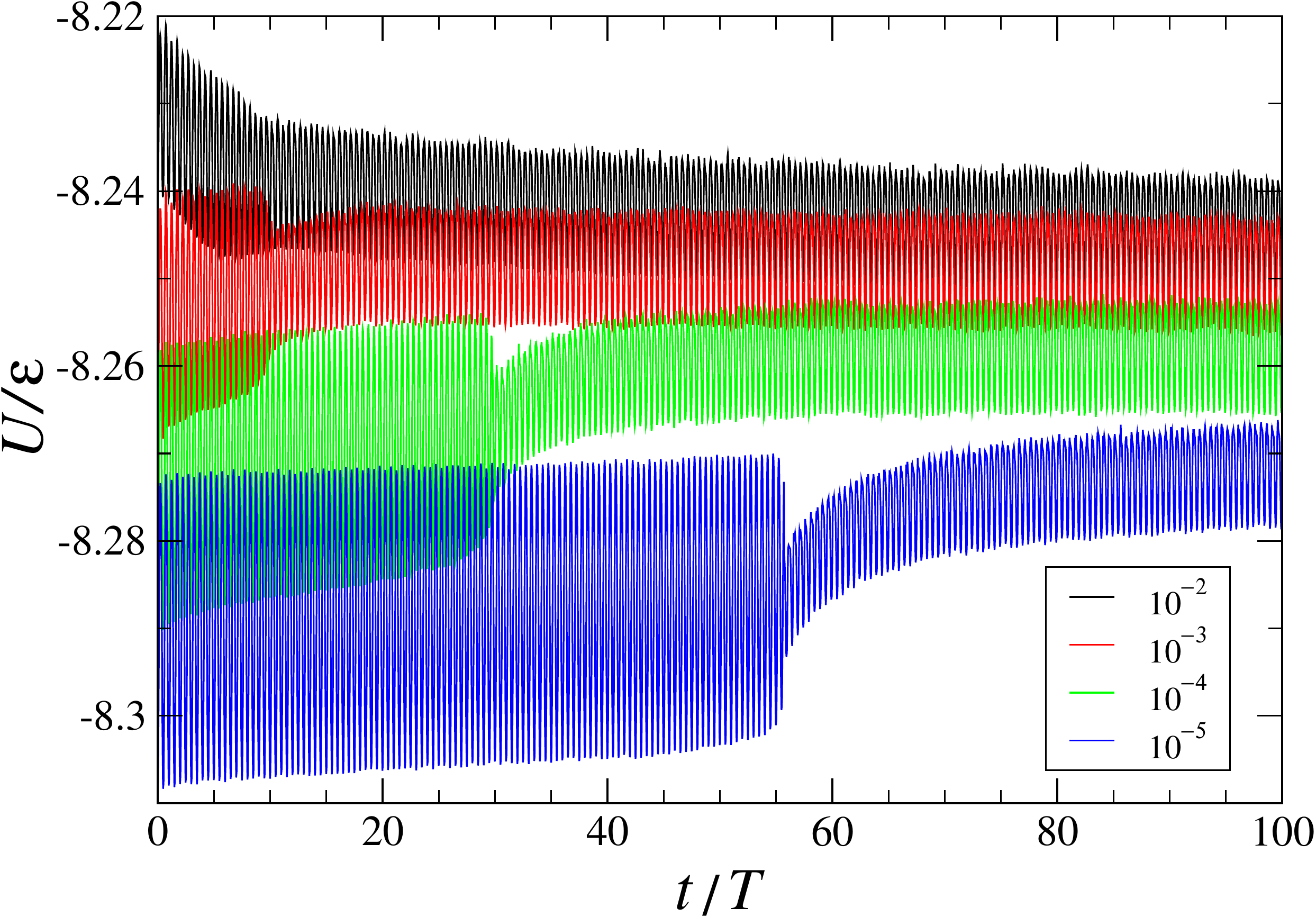}
\caption{(Color online) The potential energy series during the first
100 oscillation periods for samples prepared with the cooling rates
$10^{-2}\varepsilon/k_{B}\tau$ (black),
$10^{-3}\varepsilon/k_{B}\tau$ (red), $10^{-4}\varepsilon/k_{B}\tau$
(green), and $10^{-5}\varepsilon/k_{B}\tau$ (blue). The oscillation
period is $T=5000\,\tau$ and the strain amplitude is
$\gamma_0=0.075$.}
\label{fig:poten_2_3_4_5}
\end{figure}

The time dependence of the potential energy per atom,
$U/\varepsilon$, is presented in Fig.\,\ref{fig:poten_2_3_4_5} for
binary glasses initially prepared with the cooling rates
$10^{-2}\varepsilon/k_{B}\tau$, $10^{-3}\varepsilon/k_{B}\tau$,
$10^{-4}\varepsilon/k_{B}\tau$, and $10^{-5}\varepsilon/k_{B}\tau$.
It can be seen that with decreasing cooling rate, the glasses settle
at deeper potential energy minima, $U(t=0)$, as expected. Following
the annealing process, the oscillatory shear deformation was imposed
during 100 cycles with the strain amplitude $\gamma_0=0.075$ at the
temperature $T_{LJ}=0.01\,\varepsilon/k_B$.  In all cases, the
deformation first proceeds during a number of cycles with relatively
large amplitudes of the potential energy variation, which is
followed by the abrupt yielding transition and oscillations with a
reduced amplitude.   Except for the sample quenched with the fastest
rate $10^{-2}\varepsilon/k_{B}\tau$, the number of cycles to reach
the yielding transition increases, and the slope of the potential
energy minima versus cycle number is reduced upon decreasing cooling
rate. Further insight can be gained by examining irreversible
displacements of atoms over consecutive snapshots, as discussed
below.

\vskip 0.05in

\begin{figure}[t]
\includegraphics[width=12.0cm,angle=0]{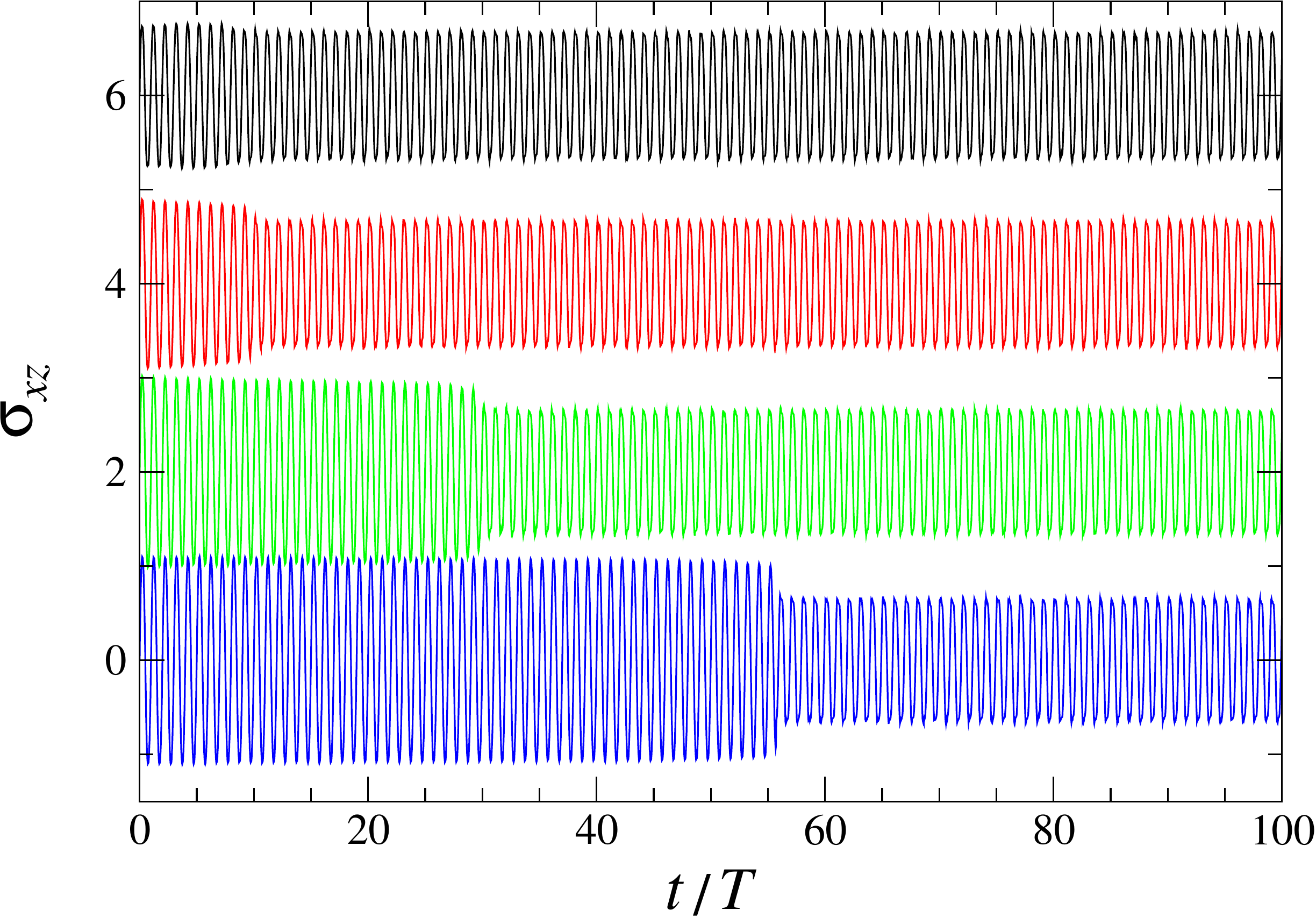}
\caption{(Color online) The time dependence of shear stress
$\sigma_{xz}$ (in units of $\varepsilon\sigma^{-3}$) during the
first 100 oscillation cycles. The glasses were initially annealed
with the rates $10^{-2}\varepsilon/k_{B}\tau$ (black),
$10^{-3}\varepsilon/k_{B}\tau$ (red), $10^{-4}\varepsilon/k_{B}\tau$
(green), and $10^{-5}\varepsilon/k_{B}\tau$ (blue). For clarity, the
data are displaced vertically by $2.0\,\varepsilon\sigma^{-3}$
(green), by $4.0\,\varepsilon\sigma^{-3}$ (red), and by
$6.0\,\varepsilon\sigma^{-3}$ (black). The strain amplitude is
$\gamma_0=0.075$ and the period of oscillations is $T=5000\,\tau$.}
\label{fig:stress_2_3_4_5}
\end{figure}
\begin{figure}[t]
\includegraphics[width=12.0cm,angle=0]{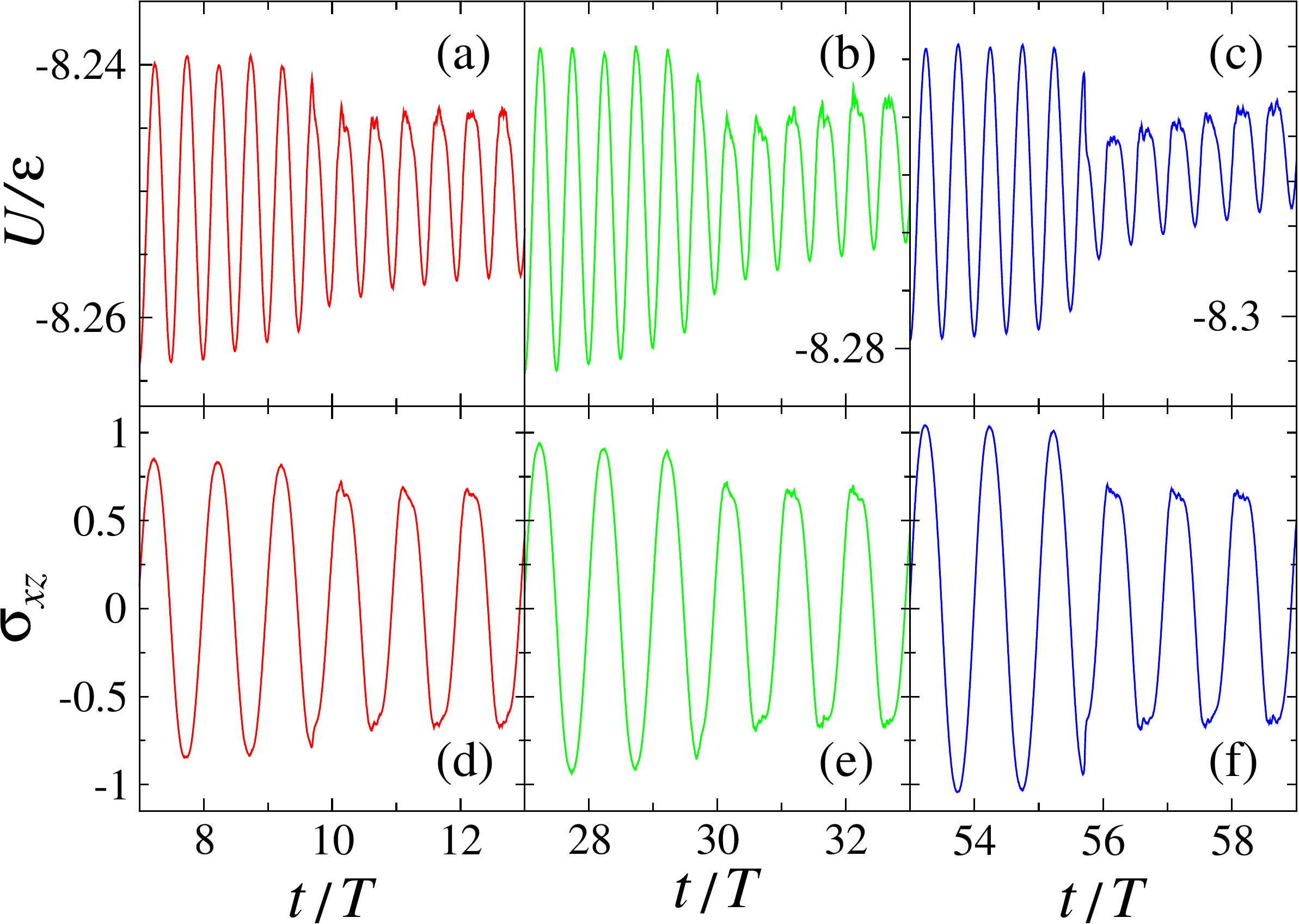}
\caption{(Color online) The enlarged view of the data for the
potential energy (a--c) and shear stress (d--f) during 6 periods
($T=5000\,\tau$) near the yielding transition. The samples were
prepared with the cooling rates $10^{-3}\varepsilon/k_{B}\tau$
(a,\,d), $10^{-4}\varepsilon/k_{B}\tau$ (b,\,e), and
$10^{-5}\varepsilon/k_{B}\tau$ (c,\,f). The same data as in
Figs.\,\ref{fig:poten_2_3_4_5} and \ref{fig:stress_2_3_4_5}.}
\label{fig:poten_stress_rem5_4_3}
\end{figure}
\begin{figure}[t]
\includegraphics[width=12.0cm,angle=0]{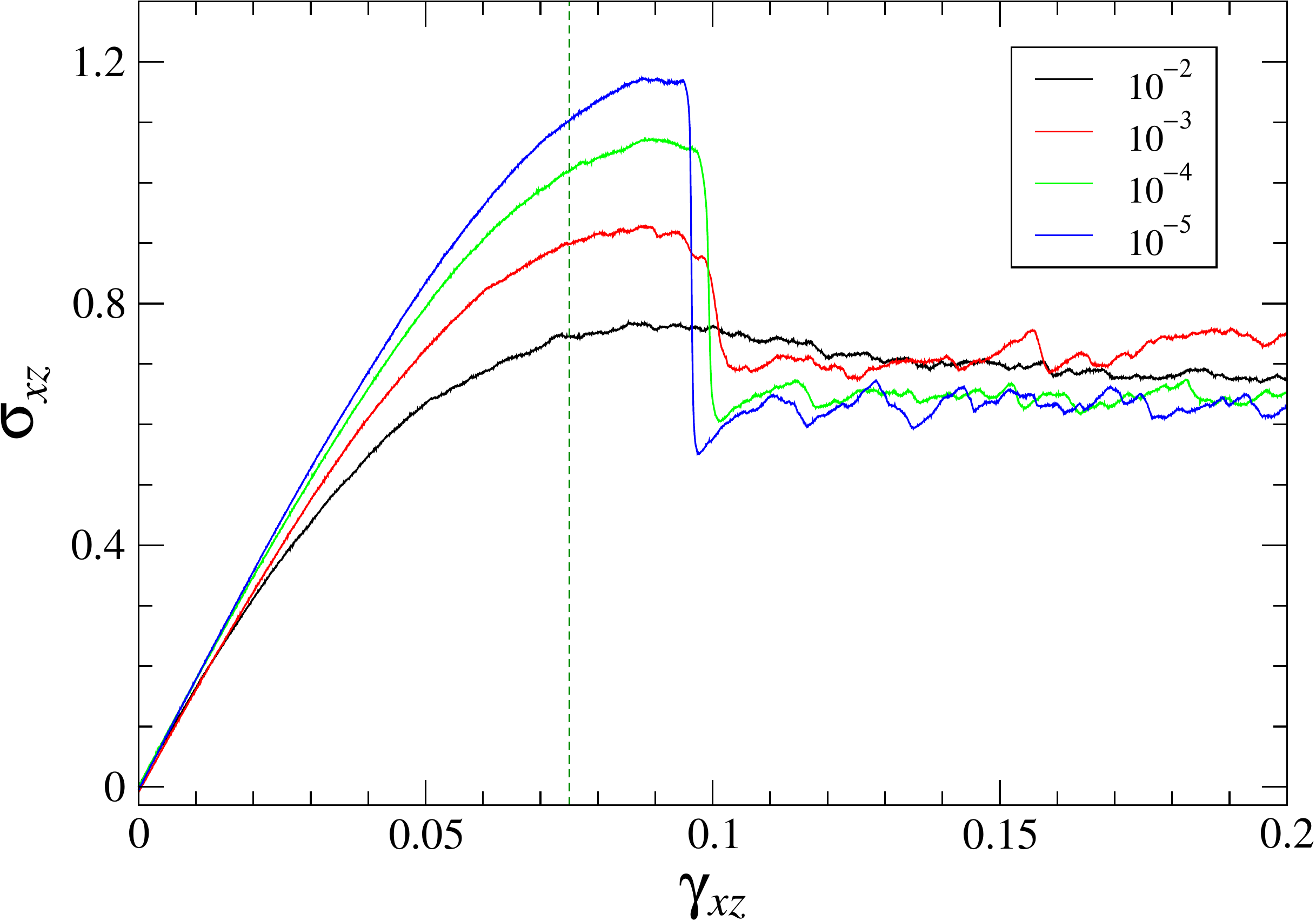}
\caption{(Color online) The shear stress $\sigma_{xz}$ (in units of
$\varepsilon\sigma^{-3}$) as a function of strain during startup
deformation with the \textit{constant strain rate}
$10^{-5}\,\tau^{-1}$ up to $\gamma_{xz}=0.20$.  The samples were
initially prepared with the cooling rates
$10^{-2}\varepsilon/k_{B}\tau$ (black),
$10^{-3}\varepsilon/k_{B}\tau$ (red), $10^{-4}\varepsilon/k_{B}\tau$
(green), and $10^{-5}\varepsilon/k_{B}\tau$ (blue). The vertical
dashed line indicates the value of shear strain
$\gamma_{xz}=0.075$.}
\label{fig:startup_stress_2_3_4_5}
\end{figure}

Next, the variation of shear stress for the same samples is
displayed in Fig.\,\ref{fig:stress_2_3_4_5} during the first 100
oscillation periods.  It can be clearly observed that, after a
certain number of cycles, the stress amplitude undergoes a
transition, typically during one period, which correlates well with
the abrupt drop in amplitude of the potential energy reported in
Fig.\,\ref{fig:poten_2_3_4_5}. The enlarged view of the potential
energy and shear stress near the yielding transition is shown in
Fig.\,\ref{fig:poten_stress_rem5_4_3}.  Note also that the stress
amplitude after 100 cycles is nearly the same in all cases, which
implies that the shear band becomes fully developed and it can
support the same maximum stress at the given strain amplitude
$\gamma_0=0.075$.  For reference, the shear stress during startup
deformation with the \textit{constant strain rate}
$10^{-5}\,\tau^{-1}$ is shown in
Fig.\,\ref{fig:startup_stress_2_3_4_5}.  Similar to periodic
loading, the glasses were also strained at constant volume and
temperature $T_{LJ}=0.01\,\varepsilon/k_B$. In particular, it can be
seen that the stress overshoot becomes increasingly pronounced for
more slowly annealed glasses; and the corresponding value of the
shear strain $\gamma_{xz}\approx0.09$ is greater than the strain
amplitude $\gamma_0=0.075$ (denoted by the vertical dashed line in
Fig.\,\ref{fig:startup_stress_2_3_4_5}) of the oscillatory
deformation.

\vskip 0.05in


A more detailed analysis of the shear band formation can be
performed by considering the spatial and temporal evolution of the
nonaffine displacements~\cite{Falk98}. Recall that the nonaffine
displacement of an atom can be evaluated via the matrix
$\mathbf{J}_i$, which linearly transforms positions of its neighbors
during the time interval $\Delta t$ and minimizes the quantity:
\begin{equation}
D^2(t, \Delta t)=\frac{1}{N_i}\sum_{j=1}^{N_i}\Big\{
\mathbf{r}_{j}(t+\Delta t)-\mathbf{r}_{i}(t+\Delta t)-\mathbf{J}_i
\big[ \mathbf{r}_{j}(t) - \mathbf{r}_{i}(t)    \big] \Big\}^2,
\label{Eq:D2min}
\end{equation}
where the sum is taken over atoms within a sphere with the radius
$1.5\,\sigma$ located at the position of the $i$-th atom
$\mathbf{r}_{i}(t)$.  This definition was first introduced by Falk
and Langer and used to identify accurately the location of shear
transformations in strained disordered solids~\cite{Falk98}. More
recently, the spatiotemporal analysis of nonaffine displacements was
applied to investigate shear band formation in
steadily~\cite{Horbach16,Pastewka19,Priez19tfic,Priez19star} and
periodically~\cite{Priezjev16,Priezjev16a,Priezjev17,
Priezjev18a,NVP18strload,PriezSHALT20} driven amorphous solids, as
well as the structural relaxation in thermally
cycled~\cite{Priez19tcyc,Priez19T2000} and elastostatically
loaded~\cite{PriezELAST19} glasses.

\vskip 0.05in

\begin{figure}[t]
\includegraphics[width=12.cm,angle=0]{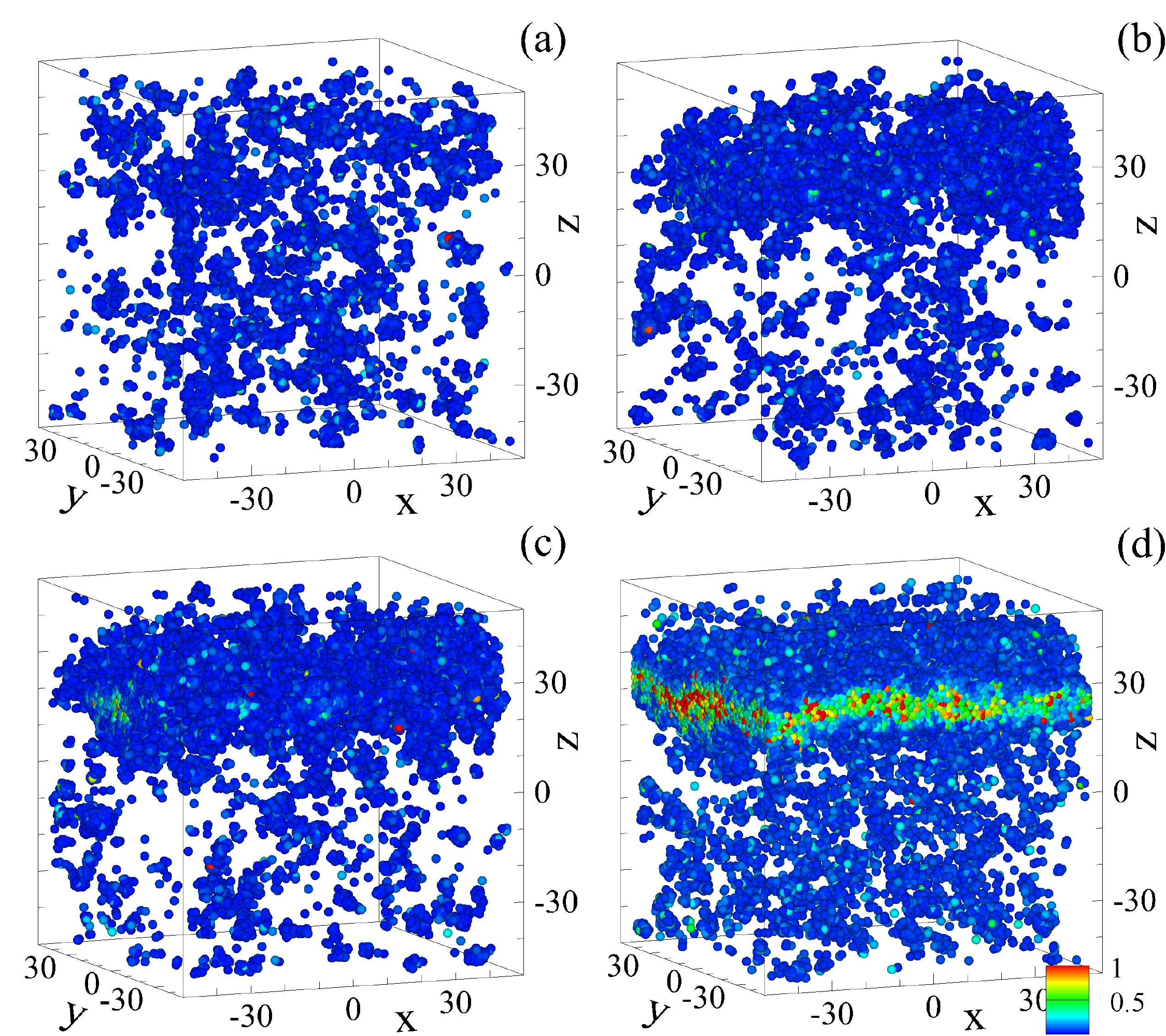}
\caption{(Color online) The snapshots of the binary glass at zero
strain after periodic deformation with the strain amplitude
$\gamma_0=0.075$ during (a) 10, (b) 50, (c) 55, and (d) 56 shear
cycles. The color denotes $D^2(nT,T)$ after a full cycle, as
indicated in the legend. The oscillation period is $T=5000\,\tau$.
Only atoms with relatively large nonaffine displacements during one
cycle, $D^2(nT,T)>0.04\,\sigma^2$, are displayed. The initial
cooling rate is $10^{-5}\varepsilon/k_{B}\tau$. The atoms are not
depicted to scale.}
\label{fig:snap_rem5_10_50_55_56}
\end{figure}
\begin{figure}[t]
\includegraphics[width=12.cm,angle=0]{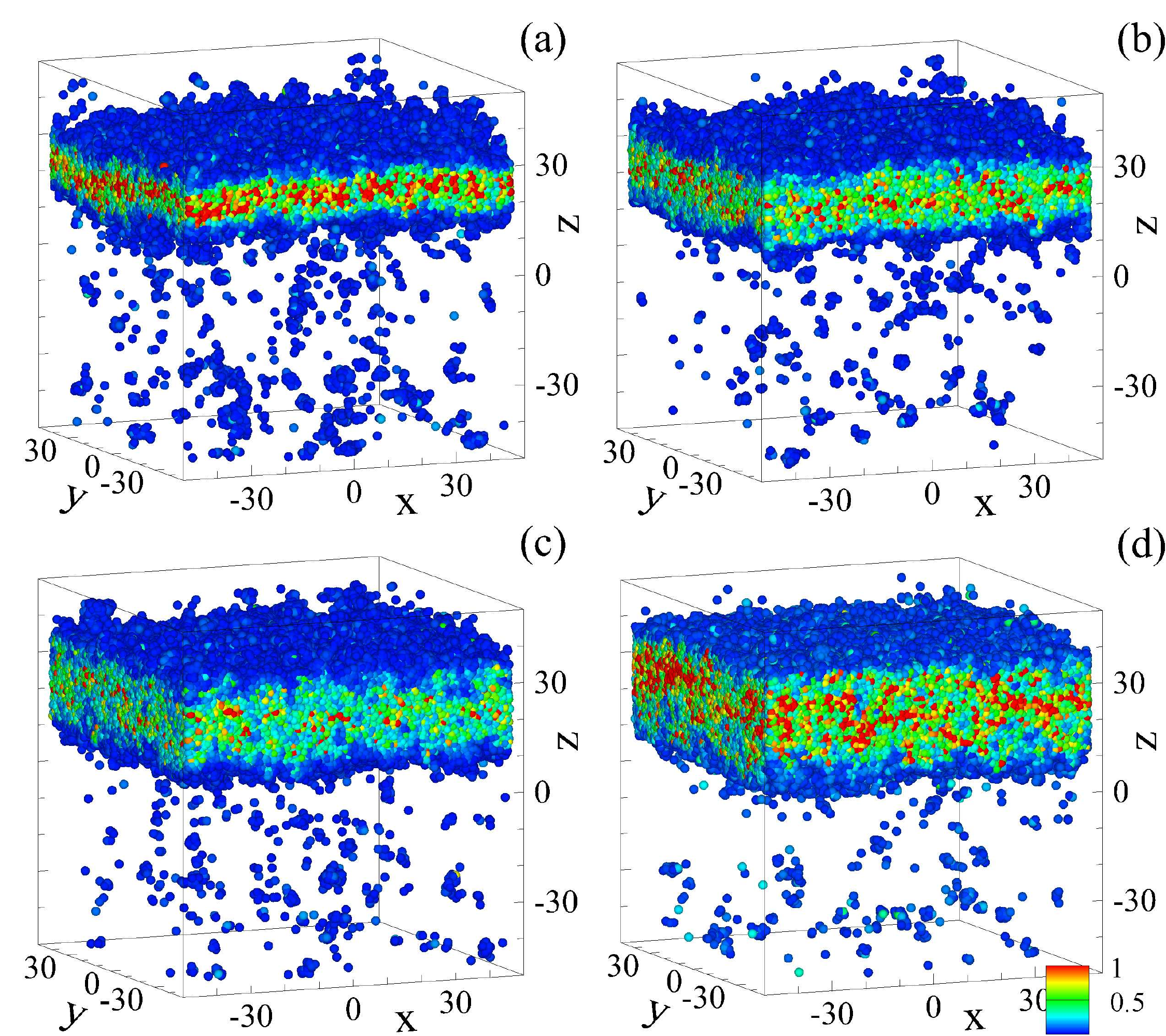}
\caption{(Color online) The atomic configurations at zero strain
after oscillatory deformation with the strain amplitude
$\gamma_0=0.075$ during (a) 57, (b) 60, (c) 70, and (d) 100 cycles.
The colorcode and the deformation protocol are the same as in
Fig.\,\ref{fig:snap_rem5_10_50_55_56}. The cooling rate is
$10^{-5}\varepsilon/k_{B}\tau$. }
\label{fig:snap_rem5_57_60_70_100}
\end{figure}

The atomic configurations and the values of the nonaffine measure
$D^2(nT,T)$ are displayed for selected cycles in
Figs.\,\ref{fig:snap_rem5_10_50_55_56} and
\ref{fig:snap_rem5_57_60_70_100} for the binary glass initially
cooled with the slowest rate $10^{-5}\varepsilon/k_{B}\tau$. In our
analysis, the quantity, $D^2(nT,T)$, was computed for any two
consecutive configurations at zero strain separated by the time
interval $\Delta t = T$.  For visualization of irreversible
displacements, the atoms with relatively small nonaffine
displacements during one cycle, $D^2(nT,T)<0.04\,\sigma^2$, are not
shown. For comparison, the typical cage size at this density is
$r_c\approx0.1\,\sigma$.  It can be clearly observed in
Fig.\,\ref{fig:snap_rem5_10_50_55_56}\,(a) that during the first 10
cycles, the mobile atoms are organized into small clusters that are
homogeneously distributed in the sample. With increasing cycle
number, the shear band starts to form at $z\approx30\,\sigma$ and it
become fully developed after 100 cycles.

\vskip 0.05in


Interestingly, the accumulation of atoms with large nonaffine
displacements along the $xy$ plane, shown in
Fig.\,\ref{fig:snap_rem5_10_50_55_56}\,(b,\,c), occurs before the
yielding transition marked by the abrupt changes in amplitudes of
$U(t)$ and $\sigma_{xz}(t)$. As is evident from
Fig.\,\ref{fig:poten_stress_rem5_4_3}\,(c,\,f), the transition takes
place during the 56-th cycle, and the corresponding atomic
configuration at the end of this cycle is shown in
Fig.\,\ref{fig:snap_rem5_10_50_55_56}\,(d). Therefore, these results
demonstrate that the location of the shear band can be predicted at
least several periods before the yielding transition.  This is in
sharp contrast to the shear band formation during the startup
deformation, where the strain localization in well-annealed glasses
becomes apparent only at the critical strain~\cite{Priez19star}.

\vskip 0.05in

\begin{figure}[t]
\includegraphics[width=12.0cm,angle=0]{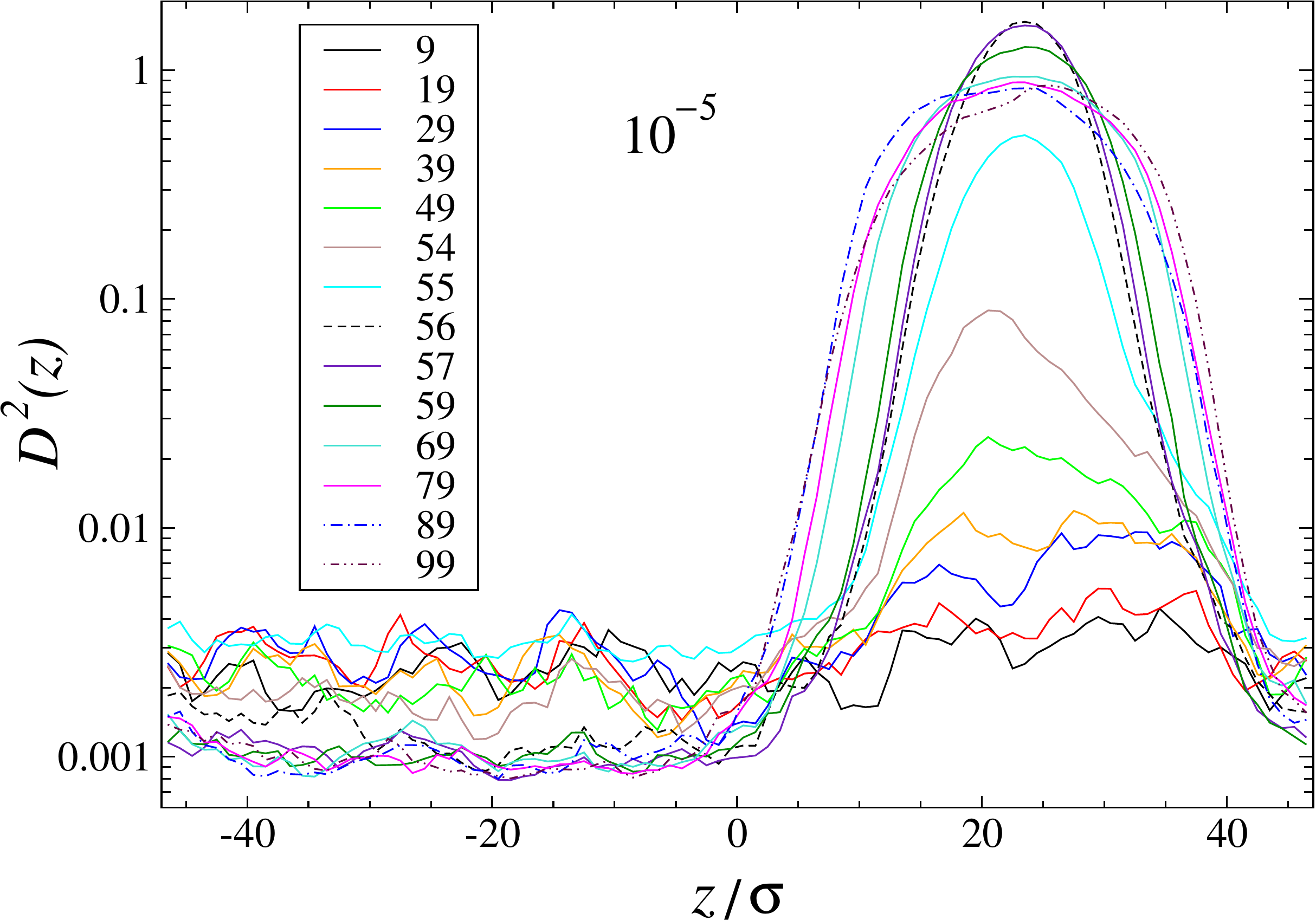}
\caption{(Color online) The variation of the nonaffine measure
$D^2(nT,T)$ for the indicated values of the cycle number, $n$. The
oscillation period is $T=5000\,\tau$. The quantity $D^2(nT,T)$ was
computed in narrow slices parallel to the $xy$ plane. The sample was
initially cooled with the rate $10^{-5}\varepsilon/k_{B}\tau$. }
\label{fig:D2min_rem5_z}
\end{figure}

Furthermore, the snapshots at larger cycle numbers shown in
Fig.\,\ref{fig:snap_rem5_57_60_70_100} reveal two effects. First,
the thickness of the shear band increases during the next 40 cycles.
Second, the typical size of clusters of mobile atoms outside the
shear band becomes smaller than before the yielding transition. This
can be explained by realizing that the amplitude of shear stress is
reduced to the maximum stress that can be supported by the shear
band, and, as a result, the solid part of the sample is strained
with a smaller amplitude.  A more quantitative description of the
spatial distribution of nonaffine displacements for the same sample
is presented in Fig.\,\ref{fig:D2min_rem5_z}.  Thus, the averaged
profiles of $D^2(nT,T)$ indicate that the location of the shear band
becomes apparent after about 30--40 shear cycles. The yielding
transition corresponds to the abrupt increase of the peak height of
$D^2(nT,T)$, from $n=54$ to $55$ in Fig.\,\ref{fig:D2min_rem5_z}.
Notice that upon further loading, the peak first becomes higher, and
then it widens to about $20\,\sigma$ at the half peak height. At the
same time, the average value of the quantity $D^2(nT,T)$ outside the
shear band is significantly reduced after the yielding transition,
which correlates well with the appearance of smaller clusters in
Fig.\,\ref{fig:snap_rem5_57_60_70_100}.  It can also be concluded
that the increase in the potential energy for $n>56$ (the blue curve
in Fig.\,\ref{fig:poten_2_3_4_5}) is associated with the increase in
volume of the shear band.

\vskip 0.05in

\begin{figure}[t]
\includegraphics[width=12.cm,angle=0]{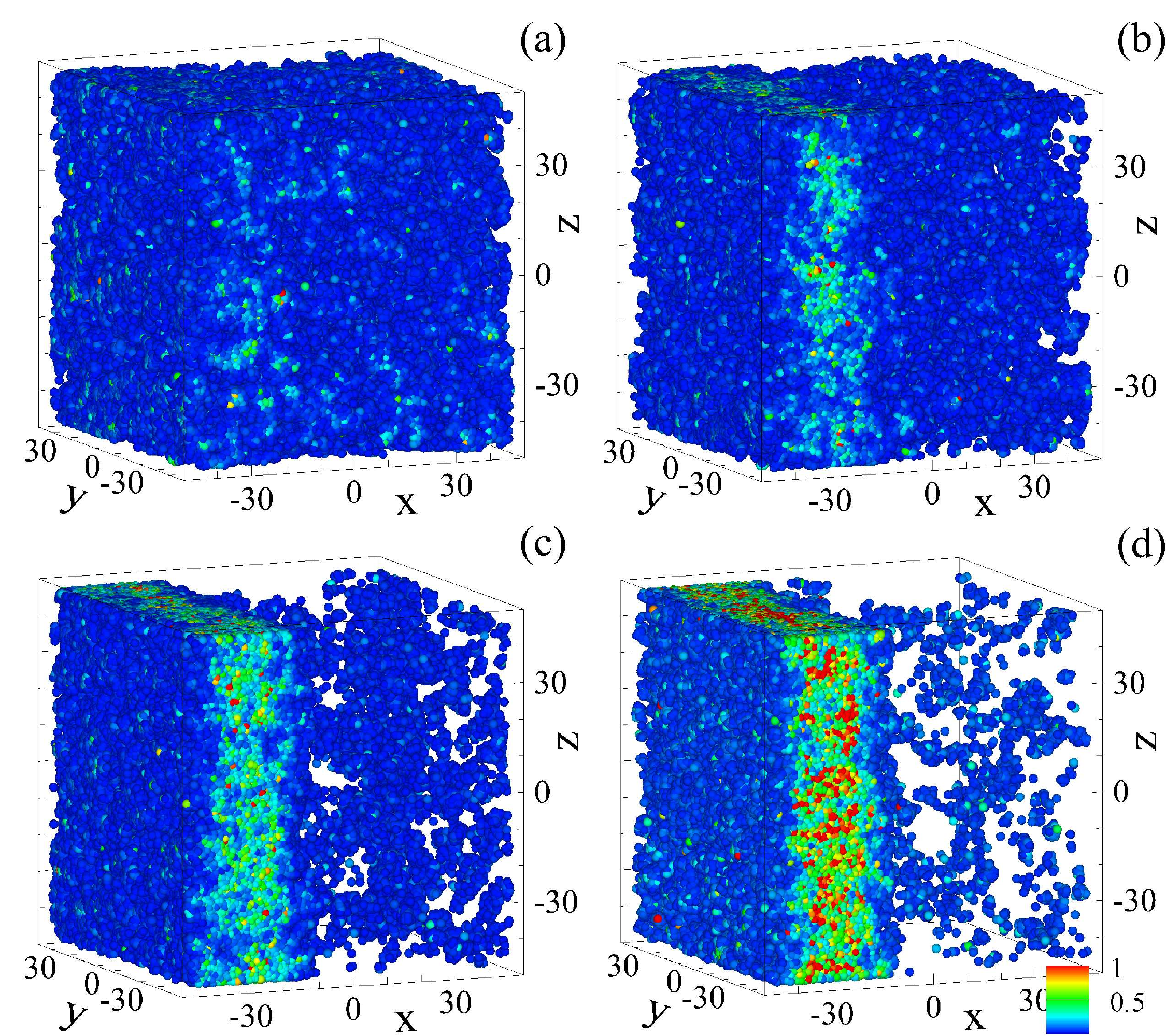}
\caption{(Color online) The configurations of atoms at zero strain
after cyclic loading with the strain amplitude $\gamma_0=0.075$
during (a) 5, (b) 9, (c) 30, and (d) 100 cycles. The colorcode for
$D^2(nT,T)$ after one cycle ($T=5000\,\tau$) is specified in the
legend. The atoms with $D^2(nT,T)<0.04\,\sigma^2$ are not shown. The
glass was cooled with the rate $10^{-2}\varepsilon/k_{B}\tau$. }
\label{fig:snap_rem2_5_9_30_100}
\end{figure}
\begin{figure}[t]
\includegraphics[width=12.0cm,angle=0]{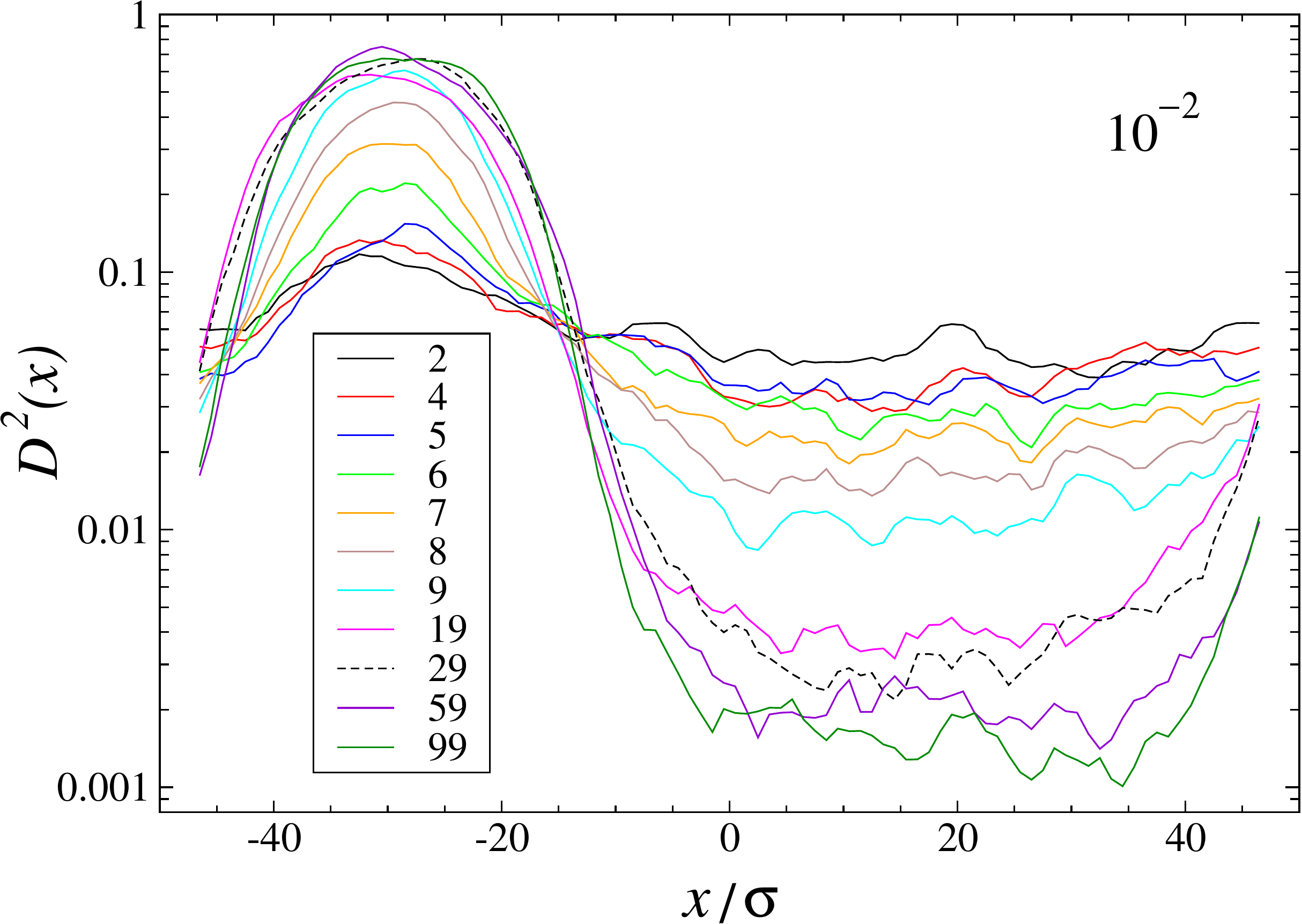}
\caption{(Color online) The nonaffine quantity $D^2(nT,T)$ as a
function of the $x$ coordinate for the indicated cycle numbers, $n$.
The oscillation period is $T=5000\,\tau$. The quantity $D^2(nT,T)$
was averaged in narrow bins parallel to the $yz$ plane. The initial
cooling rate is $10^{-2}\varepsilon/k_{B}\tau$.  }
\label{fig:D2min_rem2_x}
\end{figure}

The distribution of the nonaffine measure for the poorly annealed
glass (cooling rate $10^{-2}\varepsilon/k_{B}\tau$) is illustrated
in Fig.\,\ref{fig:snap_rem2_5_9_30_100}.  It can be observed that
initially most of the atoms undergo irreversible displacements,
followed by widening of the shear band and relaxation of the
adjacent material over consecutive cycles.  The corresponding
profiles of $D^2(nT,T)$ averaged in narrow bins along the $yz$ plane
are shown in Fig.\,\ref{fig:D2min_rem2_x}.  In this case, the
yielding transition occurs within the first few cycles, and the
shear band is formed by the 10-th cycle.  This is consistent with
the results for $U(t)$ and $\sigma_{xz}(t)$ in
Figs.\,\ref{fig:poten_2_3_4_5} and \ref{fig:stress_2_3_4_5}.

\vskip 0.05in


It can be noticed in Fig.\,\ref{fig:D2min_rem2_x} that the average
quantity $D^2(nT,T)$ outside the shear band is significantly reduced
during 100 cycles, which implies that the solid part continues to
relax to lower energy states. Therefore, the gradual decay of $U(t)$
for $n\gtrsim10$ in Fig.\,\ref{fig:poten_2_3_4_5} contains two
contributions; the first one comes from the widening of the shear
band (leading to higher energy) and the other one originates from
the mechanical annealing of the solid domain where the potential
energy is reduced. Recent studies have shown that the relaxation
process in poorly annealed glasses might continue during hundreds of
subyield cycles at a finite temperature~\cite{Priezjev18,Priezjev18,
Sastry19band,PriezSHALT20}. Thus, it can be expected that upon
further loading, the potential energy minima for the poorly annealed
glass (cooling rate $10^{-2}\varepsilon/k_{B}\tau$) will continue
approaching the energy level of the well annealed sample (cooling
rate $10^{-5}\varepsilon/k_{B}\tau$).  The long time behavior of
periodically deformed glasses, however, is not the main focus of the
present study, and, thus, it was not explored, partly due to
computational limitations.

\vskip 0.05in

\begin{figure}[t]
\includegraphics[width=12.0cm,angle=0]{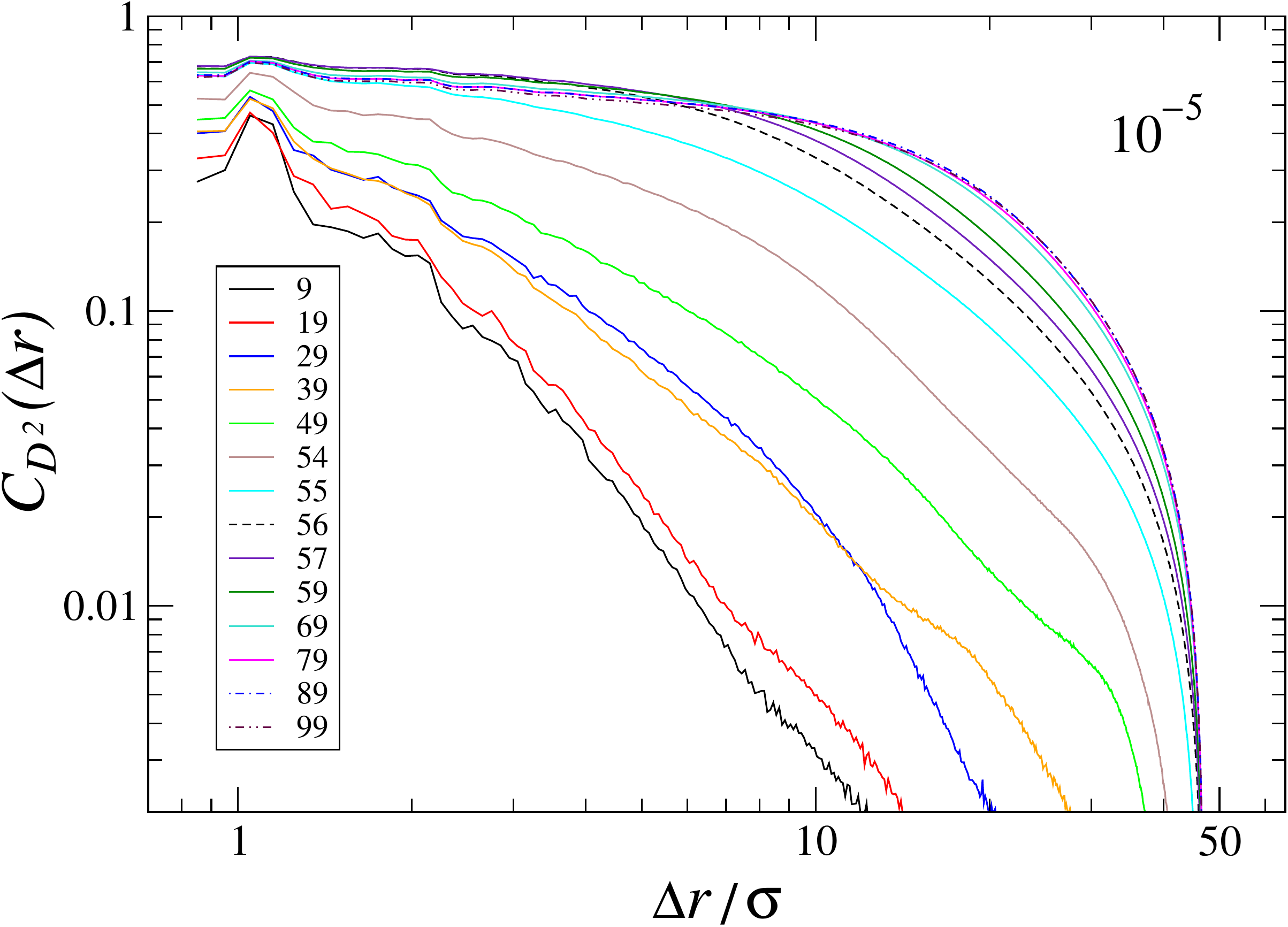}
\caption{(Color online)  The correlation function $C_{D^2}(\Delta
\textbf{r})$ defined by Eq.\,(\ref{Eq:CORR_D2}) for the sample
initially cooled with the rate $10^{-5}\varepsilon/k_{B}\tau$. The
values of the cycle number for the nonaffine measure $D^2(nT,T)$ are
listed in the legend. The same cycle numbers as in
Fig.\,\ref{fig:D2min_rem5_z}.}
\label{fig:C_D2min_rem5}
\end{figure}
\begin{figure}[t]
\includegraphics[width=12.0cm,angle=0]{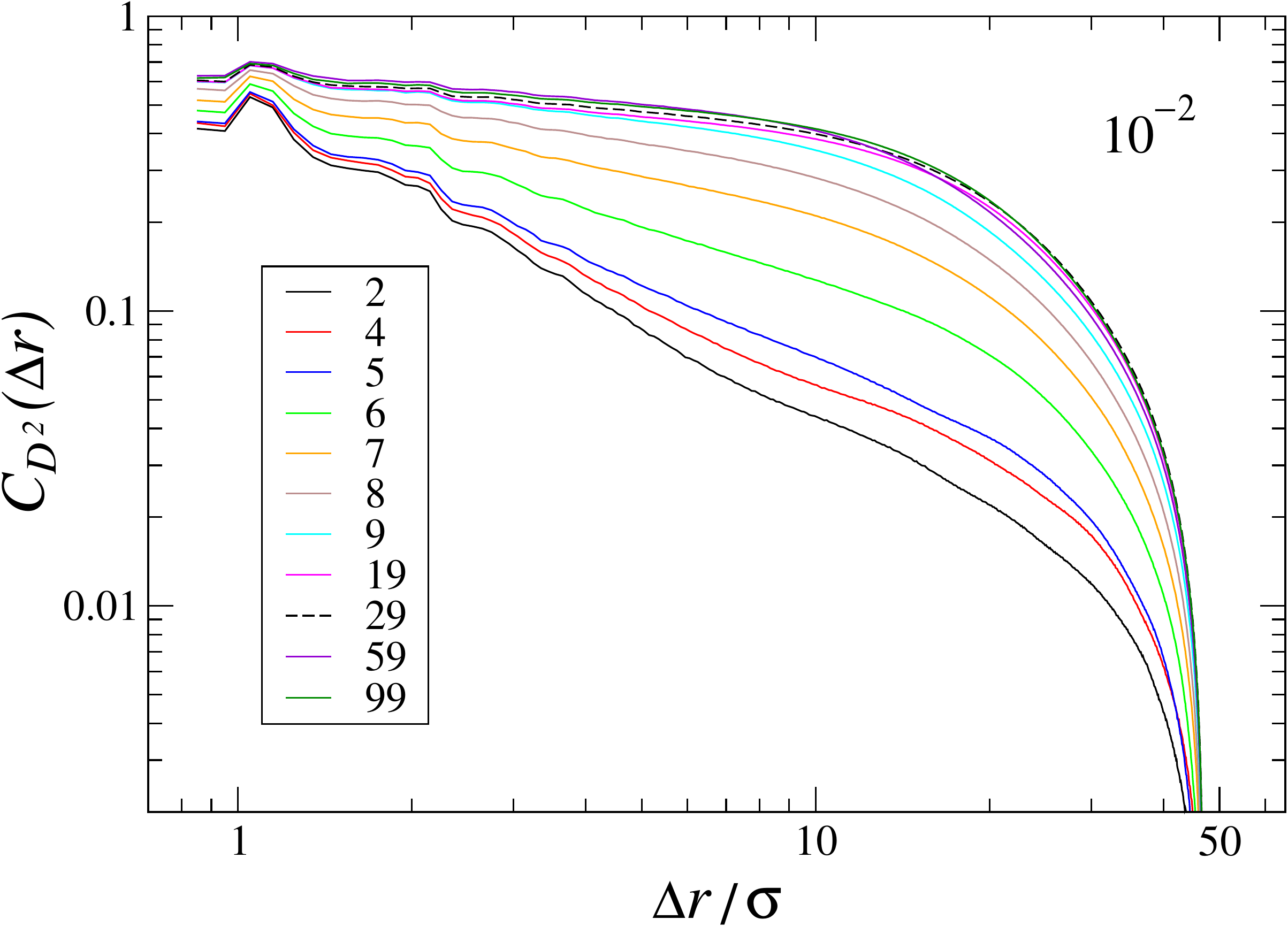}
\caption{(Color online) The spatial variation of the correlation
function $C_{D^2}(\Delta \textbf{r})$ for the indicated values of
the cycle number. Before cyclic loading, the binary glass was
annealed to the temperature $T_{LJ}=0.01\,\varepsilon/k_B$ with the
rate $10^{-2}\varepsilon/k_{B}\tau$. }
\label{fig:C_D2min_rem2}
\end{figure}

The spatial distribution of nonaffine displacements during the
loading process can be further analyzed by considering the
normalized, equal-time correlation function~\cite{Chikkadi12}, as
follows:
\begin{equation}
C_{D^2}(\Delta \textbf{r}) = \frac{\langle D^2(\textbf{r} + \Delta
\textbf{r}) D^2(\textbf{r}) \rangle - \langle D^2(\textbf{r})
\rangle^2}{\langle D^2(\textbf{r})^2 \rangle - \langle
D^2(\textbf{r}) \rangle^2},
\label{Eq:CORR_D2}
\end{equation}
where the averaging is performed over all pairs of atoms, and
$D^2(nT,T)$ is computed for two configurations at zero strain
separated by $\Delta t = T$, similar to the analysis in
Figs.\,\ref{fig:D2min_rem5_z} and \ref{fig:D2min_rem2_x}. The
results for the limiting cases of well and poorly annealed glasses
are presented in Figs.\,\ref{fig:C_D2min_rem5} and
\ref{fig:C_D2min_rem2}, respectively. It can be observed that the
function $C_{D^2}(\Delta \textbf{r})$ decays over short distances
for the cycles $n=9$ and $19$ in Fig.\,\ref{fig:C_D2min_rem5}, which
is consistent with the formation of small-size clusters in the well
annealed sample shown in Fig.\,\ref{fig:snap_rem5_10_50_55_56}\,(a).
Upon increasing cycle number, the correlation of nonaffine
displacements becomes increasingly long ranged; although the precise
cycle number for the yielding transition (detected via drop in $U$
or $\sigma_{xz}$ when $n=55$) can hardly be identified from these
data.  Similar trends are also evident in
Fig.\,\ref{fig:C_D2min_rem2} for the poorly annealed glass, except
that $C_{D^2}(\Delta \textbf{r})$ decays relatively slowly even at
small $n$ when most of the atoms undergo irreversible displacements,
as shown, for example, in Fig.\,\ref{fig:snap_rem2_5_9_30_100}\,(a).
Overall, these results indicate that during strain localization, the
dynamics of nonaffine displacements becomes correlated over larger
distances, which is reflected in the shape of the nonaffinity
correlation function.

\section{Conclusions}

In summary, the process of shear band formation and yielding
transition in disordered solids subjected to periodic shear
deformation is investigated using molecular dynamics simulations.
The binary glass consists of one million atoms, and it is formed via
annealing from the liquid state to a low temperature with different
cooling rates.  It was demonstrated that more slowly cooled glasses
undergo a transition to plastic flow after larger number of cycles
with the strain amplitude slightly above the critical value. In
contrast to startup deformation, the location of a shear band can be
predicted from the spatial distribution and averaged profiles of
nonaffine displacements at least a few cycles before the yielding
transition.  It was also shown that the collective dynamics of
nonaffine displacements during shear band formation is reflected in
the evolution of the spatial correlation function, which gradually
changes from a fast to slow decay that is nearly independent of the
cycle number when the shear band is fully developed.

\section*{Acknowledgments}

Financial support from the National Science Foundation (CNS-1531923)
and the ACS Petroleum Research Fund (60092-ND9) is gratefully
acknowledged. The numerical simulations were performed at Wright
State University's Computing Facility and the Ohio Supercomputer
Center. The molecular dynamics simulations were performed using the
LAMMPS open-source code developed at Sandia National
Laboratories~\cite{Lammps}.


%

%

%

%

%

%

%

%

%

%

%

\bibliographystyle{prsty}

\end{document}